\documentstyle[12pt]{article}
\newcommand{\beq}{\begin{equation}}
\newcommand{\eeq}{\end{equation}}
\newcommand{\beqa}{\begin{eqnarray}}
\newcommand{\eeqa}{\end{eqnarray}}
\newcommand{\ba}{\begin{array}}
\newcommand{\ea}{\end{array}}
\newcommand{\CR}{\nonumber \\}
\newcommand{\pa}{\partial}

\newcommand{\bra}{\langle}
\newcommand{\ket}{\rangle}

\begin{document}
\begin{titlepage}
\begin{flushright}
hep-th/9803126 \\
UTHEP-381 \\
March, 1998
\end{flushright}
\vspace{0.5cm}
\begin{center}
{\Large \bf 
The WDVV Equations in $N=2$ Supersymmetric Yang-Mills Theory}
\lineskip .75em
\vskip2.5cm
{\large Katsushi Ito\footnote[2]{Address after April 1, 1998: Yukawa Institute
for Theoretical Physics, Kyoto University, Kyoto 606-8502, Japan.}  
and Sung-Kil Yang}
\vskip 1.5em
{\large\it Institute of Physics, University of Tsukuba \\
Ibaraki 305-8571, Japan}
\end{center}
\vskip3cm
\begin{abstract}
We present a simple proof of the WDVV equations for the prepotential of
four-dimensional $N=2$ supersymmetric Yang-Mills theory with all $ADE$
gauge groups. According to our proof it is clearly seen
that the WDVV equations in four dimensions have their origin in the
associativity of the chiral ring in two-dimensional topological
Landau-Ginzburg models. The WDVV equations for the $BC$ gauge groups are also
studied in the Landau-Ginzburg framework.
We speculate about the topological field theoretic
interpretation of the Seiberg-Witten solution of $N=2$ Yang-Mills theory.
\end{abstract}
\end{titlepage}
\baselineskip=0.7cm

The exact low-energy effective action of $N=2$ supersymmetric Yang-Mills
theory is determined by a holomorphic prepotential, 
which is obtained from the period integrals of the meromorphic one-form,
the Seiberg-Witten (SW) differential, on a Riemann surface \cite{SeWi}.
For a simple gauge group $G$, it has been recognized that the spectral curves
of the periodic Toda lattice associated with the dual affine Lie algebra 
$(\widehat{G})^{\vee} $\cite{Go},\cite{MaWa} provide the exact solution
consistent with the microscopic instanton calculations.
In our previous papers, we obtained the Picard-Fuchs 
differential equations for the period integrals \cite{ItYa} and studied their
solutions in the weak-coupling region for the $ADE$ gauge groups \cite{IY2}. 
For our analysis, it is essential to regard the spectral curve as the 
fibration over ${\bf CP}^{1}$ whose fiber is the single-variable
version of the superpotential for the $ADE$ topological Landau-Ginzburg (LG) 
models. This relation between the four-dimensional SW theory and 
two-dimensional topological LG model suggests a hidden topological field
theoretical structure of the low-energy effective action. 
In fact, Marshakov et al. \cite{MaMiMo} have shown that the prepotential 
obeys the WDVV (Witten-Dijkgraaf-Verlinde-Verlinde) equations \cite{WDVV}, 
which play an important role in evaluating the free energy in two-dimensional
topological field theory \cite{KoMa}. 
This type of non-linear equations would provide a useful tool for
determining the instanton corrections without relying on the explicit
form of the underlying Riemann surfaces. 
They might also have an important application to investigating the
Donaldson-Witten theory for four-dimensional manifolds \cite{MoWi}.
 
The analysis of the WDVV equations in \cite{MaMiMo} is based on the 
associative property of the one-forms over the hyperelliptic Riemann surface. 
In \cite{BoMa}, on the other hand, Bonelli and Matone derive
the WDVV equation for $SU(3)$ gauge group from the Picard-Fuchs equation.
In this paper, we will obtain the WDVV equations for the $ADE$ gauge 
groups from the Gauss-Manin system that the SW period integrals
obey. It will be observed that the WDVV equations arise from those of the 
corresponding topological LG model. In a similar vein the WDVV equations for
$B_r$ as well as $C_r$ gauge groups will be obtained from the corresponding
Gauss-Manin system. 

We begin with introducing a spectral curve for a gauge group $G$ and
summarizing some basic properties of two-dimensional topological LG models.
The Toda spectral curve for a simply laced gauge group $G$ $(=A_{r},D_{r},
E_{6}, E_{7}$ or $E_{8}$) of  rank $r$ takes the form:
\begin{equation}
z+{\mu^{2}\over z}=W_{G}(x; t_{1}, \cdots, t_{r}),
\label{eq:spe}
\end{equation}
where $W_{G}(x; t_{1}, \ldots, t_{r})$ is identified as the superpotential 
for the LG models of type $G=ADE$ with flat coordinates
$t_{i}$ $(i=1,\cdots, r)$ \cite{EY},\cite{ItYa}. 
The overall degree of $W_G$ equals
$h^{\vee}$, the dual Coxeter number of $G$, and $t_{i}$ has degree 
$q_{i}=e_{i}+1$ where $e_{i}$ is the $i$-th exponent of $G$. In particular
$q_2=2$ and $q_r=h$ with $h$ being the Coxeter number of $G$. For $ADE$
gauge groups we have $h^\vee =h$.

In topological LG models it is well-known \cite{Wa} that the primary fields 
$\phi_{i}=\partial_{t_{i}}W(x)$ where 
$\partial_{t_{i}}={\partial\over \partial t_{i}}$, generate the chiral ring
\begin{equation}
\phi_{i}(x)\phi_{j}(x)=\sum_{k=1}^{r}{C_{ij}}^{k}(t)\phi_{k}(x)+Q_{ij}(x)
\partial_{x}W(x),
\end{equation} 
where ${C_{ij}}^{k}(t)$ are the structure constants.
There is a distinguished coordinate $t_{r}$ satisfying
$\phi_{r}=\partial_{t_{r}}W(x)=1$.
Note that the flatness condition implies the relation
\begin{equation}
\partial_{x}Q_{ij}(x)=\partial_{t_{i}}\partial_{t_{j}}W(x),
\end{equation} 
and the topological metric $\eta_{ij}$ is given by
\begin{equation}
\eta_{ij}\equiv\langle \phi_{r}\phi_{i}\phi_{j}\rangle
=\delta_{e_{i}+e_{j},h^{\vee}}.
\end{equation}
The associativity of the chiral ring
$\phi_{i}(\phi_{j}\phi_{k})=(\phi_{i}\phi_{j})\phi_{k}$ implies the
relation ${C_{ij}}^{l}{C_{lk}}^{m}={C_{jk}}^{l}{C_{li}}^{m}$. 
Denoting by $C_{i}$ the matrix with the elements
${(C_{i})_{j}}^{k}={C_{ij}}^{k}$, we get
\begin{equation}
\mbox{[} C_{i},\, C_{j} \mbox{]}=0.
\label{eq:wdvv1}
\end{equation}
In topological field theory there exists the free energy 
$F(t_{1}, \cdots, t_{r})$ in such a way that a three-point 
function $\langle \phi_{i}\phi_{j}\phi_{k}\rangle$ is given by 
$F_{ijk}\equiv\partial_{t_{i}}\partial_{t_{j}}\partial_{t_{k}}F(t)$. From 
the relation $F_{ijk}={C_{ij}}^{l}\eta_{lk}$ and (\ref{eq:wdvv1}),
we obtain the WDVV equations \cite{WDVV}
\begin{equation}
F_{i}\eta^{-1}F_{j}=F_{j}\eta^{-1}F_{i},
\end{equation}
where the matrix $F_{i}$ has the elements $(F_{i})_{jk}=F_{ijk}$.

Let us turn to four-dimensional $N=2$ Yang-Mills theory with the dynamical
scale $\Lambda$. The spectral curve (\ref{eq:spe}) with 
$\mu^{2}=\Lambda^{2 h^{\vee}}/4$  yields the Riemann surface
which is needed to describe the non-perturbative low-energy behavior of the
Coulomb branch. The flat directions of the Coulomb branch are parametrized by
$t_i$. The SW differential 
\begin{equation}
\lambda_{SW}={x\over 2\pi i }{dz \over z}
\label{SWdiff}
\end{equation}
is used to define the period integrals
\begin{equation}
a_{I}=\oint_{A_{I}}\lambda_{SW}, \hskip10mm 
a_{D I}=\oint_{B_{I}}\lambda_{SW}, \hskip10mm  I=1, \cdots , r
\label{periods}
\end{equation}
where $A_{I}$ and $B_{I}$ are canonical one-cycles on the curve.
The periods $\Pi=(a_{I},a_{D I})$ obey a set of differential equations,
called the Picard-Fuchs (PF) equations. They take the simple form
in terms of the flat coordinates \cite{ItYa}:
\begin{eqnarray}
{\cal L}_{0}\Pi&&\equiv
 \left(\sum_{i=1}^{r}q_{i}t_{i}\partial_{t_{i}}-1\right)^2\Pi
-4\mu^{2}(h^{\vee})^{2}\partial_{t_{r}}^{2}\Pi=0, \nonumber \\
{\cal L}_{ij}\Pi&&\equiv \partial_{t_{i}}\partial_{t_{j}}\Pi
-\sum_{k=1}^{r} {C_{ij}}^{k}(t)
\partial_{t_{k}}\partial_{t_{r}}\Pi=0.
\label{adePF}
\end{eqnarray}
The first equation arises from the scaling property of $\lambda_{SW}$
\beq
\left(\sum_{i=1}^{r}q_{i}t_{i}\partial_{t_{i}}+h^{\vee}\mu\partial_{\mu}
-1\right)\lambda_{SW}=\partial_{x}(*)\, dx.
\label{firstorder}
\eeq
The second equations are nothing but the Gauss-Manin system for the
$ADE$ singularity \cite{Gauss}. Note that the first order scaling equation 
(\ref{firstorder}) is valid for any gauge group.

Now we introduce the prepotential ${\cal F}(a)$ of the low-energy 
effective theory by 
\begin{equation}
a_{D I}={\partial {\cal F}(a)\over \partial a_{I}}.
\end{equation}
Make a change of variables from the flat coordinates $\{t_{i}\}$ to the
periods $\{a_{I}\}$, then, in terms of new variables, 
the Gauss-Manin system is expressed as
\begin{equation}
\left(\partial_{t_{i}}a_{I} \partial_{t_{j}}a_{J}
-\sum_{k=1}^{r}{C_{ij}}^{k}(t)
\partial_{t_{k}}a_{I} \partial_{t_{r}}a_{J}\right)
\partial_{a_{I}}\partial_{a_{J}}\Pi+P_{ij}^{I}\partial_{a_{I}}\Pi=0,
\label{eq:wdvv2}
\end{equation}
where
\begin{equation}
P_{ij}^{I}=\partial_{t_{i}}\partial_{t_{j}} a_{I}
-\sum_{k=1}^{r}{C_{ij}}^{k}(t) \partial_{t_{k}}\partial_{t_{r}} a_{I}.
\end{equation}
Since $a_{I}$ is the solution of the Gauss-Manin system, it is obvious that
$P_{ij}^{I}=0$ and $\Pi=a_{I}$ satisfies (\ref{eq:wdvv2}).
We next put $\Pi=a_{D I}$ in (\ref{eq:wdvv2}). This gives rise to the third 
order differential equations for ${\cal F}(a)$ of the form:
\begin{equation}
\widetilde{\cal F}_{ijk}=\sum_{l=1}^{r} {C_{ij}}^{l}\widetilde{\cal F}_{lrk},
\label{eq:wdvv3}
\end{equation}
where 
\begin{equation}
\widetilde{\cal F}_{ijk}=\partial_{t_{i}}a_{I} \partial_{t_{j}}a_{J}
\partial_{t_{k}}a_{K} {\cal F}_{IJK}, \hskip10mm
{\cal F}_{IJK}=\partial_{a_{I}}\partial_{a_{J}}\partial_{a_{K}}{\cal F}(a).
\end{equation}
Defining the metric by ${\cal G}_{ij}=\widetilde{\cal F}_{ijr}$ we have
\begin{equation}
\widetilde{\cal F}_{i}=C_{i} {\cal G},
\label{FCGeq}
\end{equation}
where $\widetilde{\cal F}_{i}$ is the matrix with entries 
$(\widetilde{\cal F}_{i})_{jk}=\widetilde{\cal F}_{ijk}$. From commutativity 
of the matrices $C_{i}$ (\ref{eq:wdvv1}) it follows that
\begin{equation}
\widetilde{\cal F}_{i} {\cal G}^{-1} \widetilde{\cal F}_{j}
= \widetilde{\cal F}_{j} {\cal G}^{-1} \widetilde{\cal F}_{i}.
\end{equation}
Thus ${\cal G}^{-1} \widetilde{\cal F}_{i}$ commute with
each other, and hence the matrices
\begin{equation}
\widetilde{\cal F}_{k}^{-1} \widetilde{\cal F}_{i}
=({\cal G}^{-1}\widetilde{\cal F}_{k})^{-1} 
{\cal G}^{-1}\widetilde{\cal F}_{i}
\end{equation}
also commute for fixed $k$. 
Therefore we obtain
\begin{equation}
\widetilde{\cal F}_{i} \widetilde{\cal F}_{k} ^{-1} \widetilde{\cal F}_{j}
= \widetilde{\cal F}_{j} \widetilde{\cal F}_{k}^{-1} \widetilde{\cal F}_{i}.
\label{eq:wdvv4}
\end{equation}
Removing the Jacobians $\partial_{t_{i}}a_{I}$ from (\ref{eq:wdvv4}) we find
the WDVV equations for ${\cal F}(a)$ 
\begin{equation}
{\cal F}_{I} {\cal F}_{K} ^{-1} {\cal F}_{J}
= {\cal F}_{J} {\cal F}_{K}^{-1} {\cal F}_{I},
\label{eq:wdvv5}
\end{equation}
where ${\cal F}_{I}$ is the matrix with $({\cal F}_I)_{JK}={\cal F}_{IJK}$.

The WDVV equations in the form (\ref{eq:wdvv5}) have been introduced 
in \cite{MaMiMo} and proved for $N=2$ Yang-Mills theory which admits the
description of the low-energy behavior in terms of hyperelliptic curves. The
associative algebra of one-differentials is employed in their 
proof.\footnote{This approach is generalized to non-hyperelliptic curves in the
case of classical gauge groups with an adjoint matter. See the second
reference of \cite{MaMiMo}.}
Our proof here applies to $N=2$ $ADE$ Yang-Mills theory for which the relevant
Riemann surfaces are not necessarily of hyperelliptic type.
It is quite clear that the WDVV equations in $N=2$ $ADE$ Yang-Mills theory 
are a consequence of the associativity of the chiral ring in two-dimensional
$ADE$ topological LG models.

Since the Gauss-Manin system for $ADE$ gauge groups does not contain the
instanton terms and $\mbox{[} {\cal L}_{ij}, 
\partial_{t_{r}} \mbox{]}=0$ \cite{IY2}, 
it is easy to work out the weak-coupling solution around $\mu^{2}=0$:
\begin{equation}
a_{I}=\bar{a}_{I}
+\sum_{k=1}^{\infty}{\mu^{2k}\over (k!)^2}\partial_{t_{r}}^{2k}\bar{a}_{I},
\end{equation}
where each coefficient in $\mu^{2}$ obeys the Gauss-Manin system. 
This means that the one-loop contribution to the prepotential
\begin{equation}
{\cal F}_{1-loop}={i\over 4\pi}\sum_{\alpha\in\Delta_{+}}(\alpha,a)^2
\log {(\alpha , a)^2\over \Lambda^{2}},
\end{equation}
where $\Delta_{+}$ denotes the set of positive roots, also satisfy the
WDVV equation (\ref{eq:wdvv5}). In this regard
it is interesting to see that the one-loop WDVV equation admits the flat
metric $K_{IJ}=(\alpha_{I},\alpha_{J})$
\begin{equation}
{\cal F}_{1-loop \ I} K ^{-1} {\cal F}_{1-loop \ J}
= {\cal F}_{1-loop \ J} K^{-1} {\cal F}_{1-loop \ I},
\label{eq:wdvv6}
\end{equation}
where $\alpha_{I}$ are the simple roots of $G$. This relation comes from
\begin{equation}
{\cal F}_{1-loop \ IJK}={i\over \pi}
\sum_{\alpha\in\Delta_{+}} {(\alpha,\alpha_{I}) (\alpha,\alpha_{J})
 (\alpha,\alpha_{K}) \over (\alpha, a)}
\end{equation}
and 
\begin{equation}
a_{I}{\cal F}_{1-loop \ IJK}={i h^{\vee}\over \pi}K_{J K}.
\end{equation}
For the $AD$ case one may check the one-loop WDVV equations (\ref{eq:wdvv6}) 
directly with the use of the explicit formula given in \cite{MaMiMo}. 
For $E_{6}$ and $E_{7}$ we have checked the above relation on the 
computer.\footnote{It is also checked numerically that the one-loop WDVV 
equations hold for $F_4$.}

So far we have studied the WDVV equations for simply laced gauge groups.
In order to proceed to the case of non-simply laced gauge groups, we wish
to write down the PF equations \cite{BCpf} in terms of the flat coordinates 
for the LG models of non-simply laced type. According to \cite{Zu}, the
non $ADE$ LG models are readily constructed as quotient of the $ADE$ models
by a discrete symmetry $\Gamma$ of the Dynkin diagram. 
Then the $BC_r$ models are obtained from the $A_{2r-1}$ models $(\Gamma =
{\bf Z}_2)$, the $F_4$ model from the $E_6$ model $(\Gamma ={\bf Z}_2)$ and
the $G_2$ model from the $D_4$ model $(\Gamma ={\bf Z}_3)$. The quotient 
procedure amounts to simply setting flat coordinates which are not 
invariant under the action of $\Gamma$ equal to zero.
Since the WDVV equation is trivial for $G_{2}$, we shall concentrate on 
the $BC$-type gauge groups in the present work.
Because of the complicated structure of the $F_{4}$ Toda spectral curve,
a detailed analysis of the $F_{4}$ gauge group is left for future study.

Let us start with the gauge group $B_{r}$. The spectral curve reads \cite{MaWa}
\begin{equation}
z+{\mu^2\over z}=\widetilde{W}_B(x;u_{1}, \cdots, u_{r}) 
\equiv {W_{BC}(x;u_{1}, \cdots, u_{r})\over x},
\end{equation}
where the LG superpotential of type $BC$
\begin{equation}
W_{BC}(x;u_{1}, \cdots, u_{r})=x^{2r}-\sum_{i=1}^{r} u_{i} x^{2r-2i}
\end{equation}
is obtained from the $A_{2r-1}$ superpotential
$W(x;\tilde{u}_{1}, \cdots, \tilde{u}_{2r-1})$ by the restriction 
$\tilde{u}_{2k}=0$
($k=1,\cdots, r-1$) and setting $u_{k}=\tilde{u}_{2k-1}$ ($k=1,\cdots, r$).
Let $\Phi_{a}$ ($a=1,\cdots, 2r-1$) be primary fields coupled to
the flat coordinates $T_{a}$ of the $A_{2r-1}$ topological LG model.
Then $t_{i}=T_{2i-1}|_{T_{2}=\cdots=T_{2r-2}=0}$  are the flat
coordinates of the $BC_{r}$ LG model. 

Now, for the SW differential 
(\ref{SWdiff}), one may follow \cite{ItYa} to derive the differential equation
\begin{equation}
\partial_{t_{i}}\partial_{t_{j}}\lambda_{SW}
=\sum_{k=1}^{r}{C_{ij}}^{k}(t)\partial_{t_{k}}\partial_{t_{r}}\lambda_{SW}
+{1\over 2\pi i}\mu\partial_{\mu}\left( {x^{-2}Q_{ij}\over 
(\widetilde{W}_B^2-4\mu^2)^{1/2}}\right)dx+\partial_{x}(*)\,dx,
\label{eq:gmb}
\end{equation}
where the total derivative term in $x$ disappears after the integration over
a closed one-cycle.
The second term in the RHS of (\ref{eq:gmb}), which is absent for
simply laced gauge groups, may be calculated as follows: From the $A_{2r-1}$
LG model one has 
\begin{equation}
Q_{ij}=\left\{
\begin{array}{cc}
0 & \mbox{for $i+j>r$,}\\
\Phi_{2(i+j)-2} & \mbox{for $i+j\leq r$.}
\end{array}
\right.
\end{equation}
Since $\Phi_{2(i+j)-2}(x)$ is an odd polynomial in $x$ and becomes 
divisible by $x$ after the restriction of the $A$ to $BC$ models we have
\begin{equation}
Q_{ij}(x)=x \sum_{k=1}^{r} {D_{ij}}^{k}(t)\phi_{k}(x),
\end{equation}
where 
\beq
{D_{ij}}^{k}(t)
={\rm res}_{\,\infty} \left( {Q_{ij}\phi_k^* \over x \pa_x W_{BC}}\right).
\eeq
Here ${\rm res}_{\,\infty}$ means to take the coefficient of $x^{-1}$ in the 
expansion at $x=\infty$ and $\phi_k^*$ is the Poincar\'e dual of $\phi_k$,
{\it i.e.} $\bra \phi_k \phi_k^* \phi_r \ket =1$.
Therefore we get the PF equation for $\Pi=\oint\lambda_{SW}$:
\begin{equation}
\partial_{t_{i}}\partial_{t_{j}}\Pi
-\sum_{k=1}^{r}{C_{ij}}^{k}(t)\partial_{t_{k}}\partial_{t_{r}}\Pi
+\sum_{k=1}^{r}{D_{ij}}^{k}(t) \mu\partial_{\mu}\partial_{t_{k}}\Pi=0.
\label{Bpf}
\end{equation}
Note that the Gauss-Manin system is not identical with that for the topological
$B_r$ LG model, but is modified in the $N=2$ $B_{r}$ Yang-Mills theory. 

As in the $ADE$ case the remaining PF equation is obtained from 
the scaling property (\ref{firstorder}) with $h^\vee =2r-1$ for $B_r$. 
Using the relation
\begin{equation}
(\mu\partial_{\mu})^2\lambda_{SW}=
4\mu^{2}\partial_{u_{r-1}}\partial_{u_{r}}\lambda_{SW} 
=-4\mu^{2}\left( t_1\pa_{t_r}^2-\pa_{t_r}\pa_{t_{r-1}} \right) \lambda_{SW},
\end{equation}
we find the scaling PF equation
\begin{equation}
\left(\sum_{i=1}^{r}q_{i}t_{i}\partial_{t_{i}}
-1\right)^{2}\Pi+4\mu^{2}(h^{\vee})^2
\left( t_1\pa_{t_r}^2-\pa_{t_r}\pa_{t_{r-1}} \right) \Pi=0.
\end{equation}

Next we discuss the gauge group $C_{r}$ for which the spectral curve 
is known to be \cite{MaWa}
\begin{equation}
z+{\mu^2\over z}=\widetilde{W}_C(x;u_{1}, \cdots, u_{r}) 
\equiv \left(x^{2}W_{BC}(x;u_{1}, \cdots, u_{r})^{2}+4\mu^{2}\right)^{1/2}.
\end{equation}
A similar procedure to the $B_{r}$ case yields the PF equation
\begin{equation}
\partial_{t_{i}}\partial_{t_{j}}\Pi
-\sum_{k=1}^{r}{C_{ij}}^{k}(t)\partial_{t_{k}}\partial_{t_{r}}\Pi
-\sum_{k=1}^{r}{D_{ij}}^{k}(t) \mu\partial_{\mu}\partial_{t_{k}}\Pi=0.
\label{Cpf}
\end{equation}
Notice that the difference between the gauge groups $B$ and $C$ is 
only the sign of the additional term in the Gauss-Manin system.
The scaling PF equation comes from (\ref{firstorder}) with $h^{\vee}=r+1$
and the relation
\begin{equation}
(\mu\partial_{\mu})^2\lambda_{SW}=
4\mu^{2}\partial_{t_{1}}\partial_{t_{r}}\lambda_{SW}.
\end{equation}
The result reads
\begin{equation}
\left(\sum_{i=1}^{r}q_{i}t_{i}\partial_{t_{i}}
-1\right)^{2}\Pi-4\mu^{2}(h^{\vee})^2\partial_{t_{1}}\partial_{t_{r}}\Pi=0.
\end{equation}

Here we wish to comment on possible instanton terms in the PF equation.
In (\ref{Bpf}) and (\ref{Cpf}), although the form of the Gauss-Manin system
deviates from (\ref{adePF}), the term carrying $\mu \pa_\mu$ is rewritten
in terms of $t_i$ variables via the scaling relation (\ref{firstorder}),
and is not the instanton effect. On the other hand, there exist
genuine instanton terms in the PF equation for the $G_{2}$ gauge
group \cite{It}. The spectral curve for $G_{2}$ is \cite{MaWa}
\begin{equation}
z+{\mu^{2}\over z}={1\over 6} (p_{1}+ \sqrt{p_{1}^2+12 p_{2}}),
\end{equation}
where 
\begin{equation}
p_{1}= 6 x^{4}-2 u x^{2}, \hskip10mm 
p_{2}= x^{8} -2 u x^{6}+u^{2} x^{4}-v x^{2} + 12 \mu^{4} .
\end{equation}
Let us introduce the flat coordinates $t_{1}=u/3$ and $t_{2}=v/6-u^3/81$. 
These are obtained by the restriction of $D_{4}$ to $G_{2}$.
The PF equation is shown to be
\begin{eqnarray}
&& \left(\partial_{t_{1}}^2-{C_{11}}^{2}\partial_{t_{2}}^2\right)\Pi
-{1\over8} t_{1} \mu\partial_{\mu}\partial_{t_{2}}\Pi 
-{8\over9} \mu^{2}\partial_{t_{2}}^2\Pi=0, \nonumber \\
&& \left(2t_{1}\partial_{t_{1}}+6t_{2}\partial_{t_{2}}-1\right)^2\Pi
-128\mu^{2}(t_{1}^2\partial_{t_{2}}+\partial_{t_{1}})\partial_{t_{2}}\Pi=0,
\end{eqnarray}
where ${C_{11}}^{2}=t_{1}^4$. Hence in this case the Gauss-Manin system
also receives instanton corrections.

We are now ready to consider the WDVV equations for the gauge groups $BC$.
It is easy to show from (\ref{Bpf}), (\ref{Cpf}) with the aid of 
(\ref{firstorder}) that the $N=2$ prepotential ${\cal F}(a)$ obeys the
third order differential equations
\beq
\widetilde{\cal F}_{ijk}-\sum_{l=1}^{r} {C_{ij}}^{l}\widetilde{\cal F}_{lrk}
-\sum_{l,n=1}^{r}b_n {D_{ij}}^{l}\widetilde{\cal F}_{lnk}=0,
\label{Fthird}
\eeq
where $b_n=\epsilon \, 2nt_n/h^\vee$ with $\epsilon =+\, (-)$ for $B_r$ 
($C_r$). In the obvious matrix notation (\ref{Fthird}) is expressed as
\beq
C_i =\widetilde C_i+\sum_{n=1}^r b_nD_i\widetilde C_n,
\label{tildeC}
\eeq
where we have set $\widetilde C_i=\widetilde{\cal F}_i {\cal G}^{-1}$ with
the same notation for $\widetilde{\cal F}$ and ${\cal G}$ as in (\ref{FCGeq}).
Suppose here that
\beq
\mbox{[}\widetilde C_i, \, \widetilde C_j \mbox{]}=0,
\label{commute}
\eeq
then the WDVV equations (\ref{eq:wdvv5}) follow immediately as seen
before. To prove (\ref{commute}) the coupled matrix-valued
equations (\ref{tildeC}) for $\widetilde C_i$ are solved explicitly since 
they are linear in $\widetilde C_i$. For instance, we find for the rank 3 case
\beqa
&& \widetilde C_1
=\pmatrix{0 & 0 & 1 \cr
	  2t_1t_2-b_3 & -b_2-t_2+t_1^2 & -b_1 \cr
	  {(\widetilde C_1)_3}^1 & {(\widetilde C_1)_3}^2
	  & b_1^2-b_1t_1-b_2},  \CR
&& \widetilde C_2=\pmatrix{0 & 1 & 0 & \cr
	      		-t_2+t_1^2 & -t_1 & 1 \cr
			2t_1t_2-b_3 & -b_2-t_2+t_1^2 & -b_1},  \hskip10mm
\widetilde C_3=\pmatrix{1 & 0 & 0 \cr
                        0 & 1 & 0 \cr
                        0 & 0 & 1 },
\eeqa
where 
\beqa
&& {(\widetilde C_1)_3}^1
=-2b_1t_1t_2+b_1b_3+t_2^2+t_1^4-b_3t_1+b_2t_2-b_2t_1^2, \CR
&& {(\widetilde C_1)_3}^2
=b_1b_2+b_1t_2-b_1t_1^2+2t_1t_2-b_3.
\eeqa
It is remarkable that these matrices indeed commute with each other.
For $B_4$ and $C_4$ we have also verified (\ref{commute}) explicitly. 
Although the WDVV equations for the $BC$ gauge groups have been proved
in a somewhat different manner \cite{MaMiMo} it will be interesting to
complete the proof along the lines presented here in the LG framework.

Finally we wish to discuss possible physical implications of the WDVV
equations in view of two-dimensional topological field theory. For this,
let us first describe an interesting observation made in \cite{BoMa2} in
which the WDVV equations (\ref{eq:wdvv5}) are extended by treating the
scale parameter $a_0 \equiv \Lambda$ on an equal footing with the
periods $a_i$. From the scaling equation for the prepotential
\begin{equation}
\left(\sum_{I=1}^{r}a_{I}\partial_{a_{I}}+a_{0}\partial_{a_{0}}\right) 
{\cal F}(a)
=2  {\cal F}(a),
\end{equation}
one obtains
\begin{eqnarray}
&& {\cal F}_{0IJ}(a)= -\sum_{K=1}^{r}a_{0}^{-1}a_{K} {\cal F}_{IJK}(a),
\hskip10mm 
{\cal F}_{00I}(a)=a_{0}^{-2}\sum_{J,K=1}^{r}a_{J}a_{K}  {\cal F}_{IJK}(a),
 \nonumber\\
&& {\cal F}_{000}(a)= -a_{0}^{-3}\sum_{I,J,K=1}^{r}a_{I}
a_{J}a_{K}  {\cal F}_{IJK}(a).
\end{eqnarray}
Define the $(r+1)\times (r+1)$ 
matrices $\widehat{\cal F}_{\alpha}$ ($\alpha=0,1,\cdots, r$) as
\begin{equation}
\widehat{\cal F}_{\alpha}=\left(
\begin{array}{c|c}
{\cal F}_{\alpha 00} & {\cal F}_{\alpha 0J} \\
\hline
{\cal F}_{\alpha I0} & {\cal F}_{\alpha IJ} \\
\end{array}
\right),
\end{equation}
then it is easy to show that the inverse of $\widehat{\cal F}_{\alpha}$ 
turns out to be
\begin{equation}
\widehat{\cal F}_{\alpha}^{-1}=\left(
\begin{array}{c|c}
A_{\alpha} & B_{\alpha} \\
\hline
{}^{t}B_{\alpha}& C_{\alpha} \\
\end{array}
\right),
\end{equation}
where 
\begin{eqnarray}
&& A_{\alpha}= {1\over 2 a_{0}^2 \sum_{J,K=1}^{r} 
a_{J} {\cal F}_{\alpha JK} a_{K}}, \hskip10mm
(B_{\alpha})_{J}= {a_{0}a_{J}\over 2  \sum_{K,L=1}^{r} a_{K}
 {\cal F}_{\alpha KL} a_{L}},
\nonumber \\
&& (C_{\alpha})_{JK}=({\cal F}_{\alpha}^{-1})_{JK}+{a_{J}a_{K}\over2 
\sum_{L,M=1}^{r} a_{L} {\cal F}_{\alpha LM} a_{M}}.
\label{eq:ext}
\end{eqnarray}
In the last equation of (\ref{eq:ext}), ${\cal F}_{0}$ is the $r \times r$
matrix given by ${\cal F}_{0}=-a_{0}^{-1}\sum_{I=1}^{r}a_{I}{\cal F}_{I}$. 
For $\widehat{\cal F}_{\alpha}$ one can now prove the extended WDVV 
equation \cite{BoMa2}
\begin{equation}
\widehat{\cal F}_{\alpha} \widehat{\cal F}_{\gamma} ^{-1}
 \widehat{\cal F}_{\beta}
= \widehat{\cal F}_{\beta} \widehat{\cal F}_{\gamma}^{-1}
 \widehat{\cal F}_{\alpha}.
\label{eq:wdvv7}
\end{equation}

This result implies that ${\cal F}(a_0,a_1,\cdots ,a_r)$ may be thought of as
the free energy of certain topological field theory whose deformations
are parametrized by the special coordinates $a_\alpha$ of special geometry.
Writing the spectral curve (\ref{eq:spe}) in the form
\beq
W_{G}^*(x,z; \mu^2, t_{1}, \cdots, t_{r})
=z+{\mu^{2}\over z}-W_{G}(x; t_{1}, \cdots, t_{r}),
\eeq
we see, as is pointed out in \cite{EY}, that $W_{G}^*$ is the sum of the 
superpotentials for the LG descriptions
of the topological ${\bf CP}^1$ model \cite{EHY} and the $ADE$ model. 
It is then natural to express $\mu^2=\mu_0^{2h^\vee}e^{t_0}/4$ where $t_0$
is the flat coordinate in the topological ${\bf CP}^1$ model \cite{Witt}.
The $ADE$ superpotential originates from a part of the defining equation
for the ALE space with complex structure deformations. On the other hand,
$t_0$ is a moduli parameter for K\"ahler deformations of the topological
$\sigma$-model on ${\bf CP}^1$ which is classified as the topological A-model.
Note that the LG description of the ${\bf CP}^1$ model is now recognized as
the mirror partner of the ${\bf CP}^1$ $\sigma$-model \cite{cp1mirror}.
Therefore it is consistent to regard $W_{G}^*$ as a superpotential for a 
topological B-model (denoted as $X_B$ henceforth) which is a tensor
product of the ${\bf CP}^1$ and $ADE$ LG models. Then it is tempting to
speculate that the SW prepotential ${\cal F}(a_0,a_1,\cdots ,a_r)$ will be the
free energy of a topological theory related to the model $X_B$. Notice that
in constructing ${\cal F}$ we have utilized a period map (\ref{periods})
\beq
\Pi :(t_0,\, t_1,\, \cdots\, ,t_r) \longrightarrow (a_0,\, a_1,\,\cdots\, ,a_r)
\label{Mmap}
\eeq
with $a_0=\Lambda =\mu_0 e^{t_0/2h^\vee}$. Note also that (\ref{eq:wdvv3}) is
written as
\beq
{\cal F}_{IJK}(a)={\pa t_i(a)\over \pa a_I}{\pa t_j(a)\over \pa a_J}
{\pa t_k(a)\over \pa a_K}\, {\cal G}_{kl}(a(t))\, {C_{ij}}^l (t(a))
\label{yukawa}
\eeq
which is reminiscent of the mirror map for the Yukawa couplings from a 
Calabi-Yau threefold to its mirror manifold. In the RHS of (\ref{yukawa})
the ``B-model Yukawa couplings'' ${C_{ij}}^l$ are free from the instanton
corrections, while in the LHS the ``A-model Yukawa couplings'' ${\cal F}_{IJK}$
receive full instanton corrections. This suggests a fascinating possibility,
though speculative, that the map (\ref{Mmap}) is understood as
a kind of mirror map
under which the prepotential ${\cal F}(a_0,a_1,\cdots ,a_r)$ is obtained as
the free energy of a topological A-model which is the mirror partner of the
model $X_B$ defined in terms of the superpotential $W_{G}^*$.
It may be worth pursuing further the idea discussed here toward uncovering
the hidden topological field theoretic property of the SW solution of $N=2$
Yang-Mills theory. We expect that the $M$-theory/Type II fivebrane 
interpretation of the SW solution will certainly play a role \cite{Mfive}.

\vskip10mm

The work of S.K.Y. was supported in part by the
Grant-in-Aid for Scientific Research on Priority Area
``Physics of CP violation'', the Ministry of Education, Science and Culture,
Japan.

\newpage


\end{document}